\documentclass[%
 reprint,
 amsmath,amssymb,
 aps,
]{revtex4-1}
\usepackage{array}
\usepackage{pbox}
\newcolumntype{L}[1]{>{\raggedright\arraybackslash}p{#1}}
\usepackage{graphicx}
\usepackage{dcolumn}
\usepackage{tikz}
\usepackage{bm}
\usepackage[toc,page]{appendix}
\usetikzlibrary{arrows.meta}
\usepackage{multirow}
\usepackage[capitalize]{cleveref}
\usepackage{setspace}
\makeatletter
\gdef\@ptsize{0} 
\makeatother

\usepackage{setspace}
\makeatletter
\gdef\@ptsize{0} 
\makeatother

\usepackage{xargs}                      

\begin{document}


\preprint{APS/123-QED}

\title{A Leaky Pipe Dream?
A Study of Gender Differences in Undergraduate Physics}

\author{Steven Turnbull}
 \affiliation{Faculty of Education and Social Work, University of Auckland, New Zealand.}
 \affiliation{Te P\={u}naha Matatini}

\author{Dion R. J. O'Neale}
\affiliation{Department of Physics, University of Auckland, New Zealand.}
\affiliation{Te P\={u}naha Matatini}%
\author{Frederique Vanholsbeeck}
\affiliation{Department of Physics, University of Auckland, New Zealand.}
\affiliation{The Dodd-Walls Centre}

\author{S. Earl Irving}
\affiliation{Faculty of Education and Social Work, University of Auckland, New Zealand.
}

\author{Thomas Lumley}
\affiliation{Department of Statistics, University of Auckland, Auckland, New Zealand.
}%

\date{\today}

\begin{abstract}
Students face diverse pathways as they journey through undergraduate study. The analysis of student course records can untangle common patterns in course progression, and identify group trends in student outcomes. 
The current work examines the relationship between gender and undergraduate physics study, using course records from over nine thousand students who enrolled in physics at the University of Auckland, spanning a six year period.

Physics students' demographic and course records were analyzed to find out whether there were gender differences in subject selection, course performance, and confidence. 
Subsequent to taking a first year physics course, female students were more likely to take further courses in life science subjects, while male students were more likely to take physical science subjects. 
In first year courses, gender differences were not present among highly academically prepared students, for whom school type (single-sex or coeducational) was a better predictor of course outcome. However, of those students who were less academically prepared in their first year, male students tended to outperform female students. Female students were also more likely to take an introductory physics course before an advancing course, compared to male students, after controlling for academic preparation.
Science capital\citep{Archer_2014}, a concept related to Pierre Bourdieu's notions of capital and habitus, was employed as an interpretive research framework. Habitus, the system of dispositions one uses to interpret the world, is largely influenced by the socio-cultural context in which an individual builds their identity. The following study explains how an interaction between science capital and an individual's habitus may lead to gender disparities in student outcomes in the field of physics.
Taken together, our findings suggests that physics may not be seen as a domain that is welcoming to female students, and the science capital framework allows us to understand why this may be the case. Although our study outlines findings from one particular sample of students, the methodology and research framework has universal applications.

\end{abstract}

\pacs{Valid PACS appear here}
\maketitle


\section{Introduction}
\label{intro}
Historically, women have been underrepresented in Science, Technology, Engineering and Mathematics (STEM) fields (especially physics). This is a concerning issue today in many different countries, and at all stages of post-compulsory education \citep{Abraham_2014,Stevanovic_2013,Smith_2011}. Much research has been dedicated to the understanding the extent, causes, and possible solutions to this problem \citep{Brewe_2016, Blickenstaff_2005}. For example, a recent special edition of the Physical Review Physics Education Research journal focused on highlighting research articles dedicated to the underrepresentation of women in physics \citep{Brewe_2016}, while there is increasing support for initiatives to advance gender equality in STEM subjects, such as Athena SWAN \citep{AthenaSwan} and Project Juno \citep{ProjectJuno}. 

It is important to understand the motivations behind gender equity research. Ensuring gender parity in STEM provides benefits, not only to women in these fields, but also society as a whole. Ensuring that all students perceive STEM subjects as viable educational paths can maximize the number of skilled workers in STEM fields, which drives economic growth. Increasing the number of women in STEM also means that more diverse perspectives are brought to research \citep{Rosser_2012,Milgram_2011}. For students who do not seek careers in STEM, choosing to take elective educational courses in STEM subjects will improve scientific literacy, which plays an important role in life outcomes \citep{Archer_2015, Irwin_2001}. Beyond the individual and societal benefits of gender equity, we, as educators and researchers, should be primarily motivated by the egalitarian ideal that every student should have a fair chance of success in their chosen field, and not be disadvantaged by factors relating to gender.

Current STEM enrollment patterns often show that female students continue to be under-represented in the physical sciences (physics, engineering, mathematics)\citep{NSF}. In contrast, this does not seem be the case in the life science subjects (biology, medicine). Why do we see gender differences in the physical science subjects, but not the life science subjects? Seeking an answer to this question, the current study used quantitative analysis to investigate the outcomes for male and female physics students at the University of Auckland - the largest university in New Zealand. Of specific interest were the gender differences in physics students' later subject selection, course performance, and confidence. We interpret our results using a theoretical framework based on the work of Pierre Bourdieu. Through this framework, we are able to relate our findings to a broader socio-cultural context.

\section{Theoretical Framework}
The metaphor of the leaky pipeline is often used to describe the attrition of women from physics\citep{Schiebinger_2001}, in that women are more likely to drop out with each transition between key stages of education (e.g., high school to university). This metaphor can be criticized for not only stigmatizing individuals that drop out of the pipeline, but for also being too simplistic \citep{Cannady_2014}. It is important to emphasize contextual factors, such as the presence of gender-stereotypes \citep{Nosek_2009} that socialize individuals to conform to gender roles (e.g., men do science, women do art). It is also important to consider the complex nature of students' subject enrollment patterns; in reality a student's journey through university study follows a complex network of unique pipes, rather than a singular pipeline. In order to understand our results in a wider context, the current study employed a science capital framework \citep{Archer_2015}, adapted from the work of Pierre Bourdieu.

The idea of capital can be used to inform gender equity research in science \citep{Kelly_1985}. Originally conceived within economics, capital has been defined as ``That part [of a man's wealth]
that he expects to provide him with \textellipsis income\textellipsis'' \cite[p.214]{Smith_1887}. \citet{Bourdieu_1986} interpreted capital as a legitimate, valuable and exchangeable resource that individuals can use to gain advantage in society. According to \citet{Bourdieu_1986}, capital has four forms: economic (e.g., financial resources), cultural (non-financial assets, such as physical appearance or style of speech), social (e.g., an individual's social network), and symbolic (prestige and recognition, such as awards). Individuals who have higher levels of capital are more able to translate their capital into personal and social advantage, such as better paid employment or education. Institutionalized education offers a mechanism by which individuals can convert their existing forms of capital into educational capital (e.g., qualifications). However, individuals who do not share the same cultural capital as the institution find themselves disadvantaged by a ``hidden curriculum''\citep{Kidman_2013}. The hidden curriculum is a term used to describe the manner in which educational institutions implicitly assess a student's ability to learn cultural norms and values, not necessarily associated with any explicit curriculum. In simple terms: a student can gain educational advantage by speaking, writing or presenting themself in a way that is favored by the educational institution. 

With regards to science education, Bourdieu does mention the idea of a science capital \citep[p.82]{Bourdieu_1985}, but his description is limited. \citet{Archer_2014} expanded on Bourdieu's work by using the term science capital to refer to the cultural and social forms of capital that are associated with science. Science capital can be used as ``a lens for understanding uneven patterns in science participation'' \citep[p.923]{Archer_2015}. Social forms of science capital may include knowing people who hold science qualifications, while cultural forms of science capital include science-related behaviors (e.g., watching science-related television shows, visiting science museums etc.). Capital also interacts with habitus; a term defined by Bourdieu as a ``system of dispositions''\citep[p.170]{Bourdieu_1985}. An individual's habitus is formed, to a great extent implicitly, in relation to the socio-cultural context of a field. For example, male and female students' habitus will be differentially impacted by the societal stereotype that science is a male-domain\citep{Nosek_2009}. As a result, female students may be more likely to internalize a belief that physics is not what women do, and they should seek a career path that is congruent with societal expectations. Habitus generates an individual's lifestyle in a field by influencing both their appreciation of the field, and their practices (behaviors) in the field. According to Bourdieu, the collection of each individual lifestyle produced by habitus then constitutes the ``represented social world''\citep{Bourdieu_1985}. 

In any given field, individuals who have similar levels of capital, or share similar socio-cultural contexts, will have similar habitus. This, in turn, produces similar practices in the field \citep[p.101]{Bourdieu_1985}. For example, a student with high levels of science capital, in addition to a habitus that predisposes them to science, will be more likely to pursue a science career. The interactions between the different forms of science capital, habitus and field of study thus contribute to an individual's science identity and performance (see Figure \ref{fig:Model}), and offer an explanation for why we often see disparities in subject enrollments between different groups of students.

The translation of capital into educational advantage may be facilitated for individuals who share characteristics with the dominant cultural group in a field, and hindered for individuals who do not. Thus, women may find themselves disadvantaged in certain STEM subjects because they do not represent the dominant group. Women may also choose not to pursue certain STEM subjects due to the role that habitus plays in influencing their perceptions of what subjects they should be doing. This idea was previously outlined by Bourdieu: ``one sees the effect of the dispositions associated with gender which help to determine the logic of\textellipsis the choice of the type of educational capital, more often literary for girls, more often scientific for boys.'' \cite[p.105]{Bourdieu_1985}.
\section{Background}
\label{sec:1}
Research consistently shows that female students tend to be underrepresented in the physical science subjects, while this is not the case in life science subjects. This pattern has been observed at high school level in Australia \citep{Kennedy_2014}, New Zealand \citep{EducationCounts_2016b}, the United Kingdom \citep{InstituteofPhysics_2012, Sheldrake_2015}, France \citep{Stevanovic_2013} and the United States \citep{Cunningham_2015,KostSmith_2010}, and university level in the United Kingdom \citep{Smith_2011}, United States \citep{Heilbronner_2012} and Ethiopia \citep{Semela_2010}. 

In the United States, only around 20\% of students studying physics at bachelors, masters or doctorate level in 2014 were female  \citep{NSF}. This contrasts with bioscience, where around 50-60\% of students studying at bachelors, masters or doctorate level were female \citep{NSF}.

Analysis of data from New Zealand in 2015 shows that science had a balanced gender-ratio of year 13 students \citep{EducationCounts_2016a}. Despite this, male students dominate the physical science subjects during year 11 at high school, with this trend continuing during later high school and university \citep{EducationCounts_2016a}. On the other hand, female students tend to be over-represented in biology and human anatomy. These gender differences are maintained at university and post-graduate levels \citep{EducationCounts_2016a}. Other data from New Zealand \citep{EducationCounts_2016b} shows that approximately 25\% of students specializing in physics and astronomy at bachelor degree level in 2015 were female, while female students tended to be over-represented in biology and health (65\% and 76\% respectively). With regards to doctoral students, approximately 29\% of students in Physics and Astronomy were female, while female students comprised 52\% of doctoral students studying biology and 68\% of those studying medicine. 

\subsection{Differences in Intentions}
\label{sec:2}
Gender disparities in STEM education can be interpreted in relation to Bourdieu's notion of habitus. As outlined by \citet{Bourdieu_1985}, an individual's habitus influences their perception of the world, whilst also being influenced by an individual's socio-cultural context at the same time. \citet[p.885]{Archer_2012} explain this idea further: ``social axes of `race'/ethnicity, social class, and gender all contribute to shaping what an individual perceives to be possible and desirable.'' An individual's habitus may be similar to another individual when they share common characteristics (e.g., gender). Individuals with similar habitus may be predisposed into behaving in similar ways \citep[p.434]{Reay_2004}. This may explain why gender differences STEM subject interest are so common. 

A meta-analysis by \citet{Su_2009} found that men tend to be more interested in engineering, science and mathematics careers compared to women, and this finding has since been replicated in a Norwegian study of secondary and tertiary physics students \citep{B_E_2013}. Data from the United States shows that high school graduates are more likely to like science and mathematics if they are male \citep{Cunningham_2015}. The same data set also shows that female students are more likely to earn credits in biology, chemistry and health subjects, while male students are more likely to earn credits in physics and engineering subjects. Similarily, cross-cultural, international data from the Programme for International Student Assessment (PISA) in 2006 has shown that among high achieving students, female students were more likely to be interested in medicine, while male students were interested in physics and chemistry \citep{Buccheri_2011}.

Further evidence suggests that the gender differences in STEM subject interest may not be present in early childhood, but emerge by the end of high school \citep{Baram_Tsabari_2010}. These stereotypical gender preferences may also persist when it comes to the types of science-related career that high school students aim for \citep{Kj_rnsli_2011}, and students' choice of STEM major at university \citep{Bottia_2015, Sadler_2012}. At university level, gender disparities may even widen further; a study of physics students at a United States university found that female students are more likely to see their interest in physics diminish during introductory physics \citep{KostSmith_2010}.

\subsection{Performance in STEM} 
\label{sec:3}

A common explanation for the gender disparities in STEM interest is the idea that female students do not perform as well as male students in STEM \citep{Blickenstaff_2005}. However, current research shows mixed findings in this area. The 2011 Trends in International Mathematics and Science Study (TIMSS) compared the science performance of male and female students in the 4th and 8th grade (years 5 and 9 in New Zealand) across 50 countries \citep{Mullis_2012}, finding much variation between countries in the level of gender differences. In the 4th grade, 23 of the 50 participating countries (including New Zealand) showed no significant gender differences, although 16 showed differences favoring boys \citep{Martin_2012}. By 8th grade, overall, female students tended to outperform male students in science. Despite this, in New Zealand, boys tended to outperform girls. Results from both the 2012 and 2015 PISA surveys found that, in New Zealand, boys have had higher mathematics and science scores at age 15 \citep{May_2013, May_2016}. 

At more advanced levels, there are also mixed findings regarding gender differences in physics performance. In the United Kingdom, studies tend to find no notable differences in the performances of male and female secondary students in mathematics \citep{Sheldrake_2015} and the sciences \citep{Smith_2011}, and this has remained relatively stable since the 1960s \citep{Smith_2011}. Similarly, \citet{Sharma_2011} found no significant gender differences in introductory physics performance at an Australian university. However, studies of university physics students have found gender differences favoring male students in concept inventories scores \citep{Pollock_2007,Kost_2009}. In a study of physics students from one university in the United States, \citet{KostSmith_2010} found that, despite male and female students performing about equally on a conceptual pre-test, male students outperformed female students on a conceptual post-test. The same study found that female students tended to outperform female students on homework tasks, but male students outperformed female students in exams. \citet{Madsen_2013} reviewed studies that have utilized concept inventories to measure the gender gap in physics at university level. They found that, overall, men tend to outperform women on pre-test concept inventories, and post-test concept inventories (although there is greater variation in the gender differences in post-test inventory scores). 

Through the lens of science capital theory, it could be argued that the male embodiment of science capital facilitates the acquisition of identity in the physical sciences, but does not necessarily translate into an increased academic advantage. This argument is supported by studies indicating that other factors beyond gender contribute to the gender gap in university level physics \citep{Kost_2009}. Furthermore, female physics students tend to be more likely to drop out of key physics courses at university, even after controlling for academic factors such as student preparedness \citep{Ellis_2016, Ost_2010}. One study failed to replicate these findings for female students in the life sciences \citep{Ost_2010}, which demonstrates how the field of study moderates the relationship between gender and the conversion of science capital into persistence. The above findings point to alternative explanations for the gender disparities seen in STEM representation that can be linked to a student's habitus.

\subsection{The Impact of Society, the Family and Career Prospects}
\label{sec:4}

As outlined by \citet{Madsen_2013}, the gender gap in physics education is likely the result of a combination of many factors, rather than a single one. This idea is summed by \citet{Blickenstaff_2005}, who suggested that women's underrepresentation in STEM points to a gender filter. The gender filter contains many layers, each one a possible reason for why women may be less likely to progress in physics. Every student holds a belief about what is a realistic educational path. However, these beliefs are informed, implicitly and explicitly, by evidence in the environment. How are people like me treated in this subject? Do my family value this subject? How many people like me study this subject? The answers to these questions will either boost or reduce a student's disposition towards a subject. The following section outlines several factors present in the socio-cultural context of physics in which habitus is formed. We will discuss some of the influencing factors located in three interconnected contexts: society (i.e., stereotypes, gender roles), the family, and careers prospects in physics. 

Many factors contributing to the gender filter are firmly rooted in society's representation of science --- especially concerning negative gender stereotypes. There are many stereotypes that target women in STEM (e.g., the belief that women are bad at mathematics), and they can negatively affect how female STEM students are viewed. Regardless of gender, undergraduate STEM students are more likely to attribute a female student's failure to internal factors (e.g., lack of ability), and a male student's failures to external factors (e.g., bad luck)\citep{LaCosse_2016}. Negative gender stereotypes persist after university; women seeking careers in STEM may be perceived as less competent by their peers \citep{Moss_2012}. Implicit gender biases relating to women in STEM are pervasive and consistent across countries \citep{Nosek_2009}. Furthermore, countries with higher levels of implicit bias tend to show greater gender disparities in 8th grade science, favoring male students \citep{Nosek_2009}. This may be due to the fact that male students' self-concept in science may be enhanced when they perceive that adults hold gender stereotypes favoring men over women \citep{Beth_Kurtz_Costes_2008}.

Women may also experience stereotype threat. Stereotype threat is a psychological construct, first described by \citet{Steele_1995}, that occurs when an individual believes that they are in a situation where a negative stereotype can be applied to them. When they feel at risk of confirming the stereotype, performance may be hindered. Female students in STEM may be discriminated against because of their gender \citep{Barthelemy_2016}, which can increase the desire to change major \citep{Steele_2002}. Stereotype threat can negatively impact on a student's career choice in STEM, particularly in male-dominated subjects like physics \citep{Deemer_2014}. \citet{Marchand_2013} found that nullifying stereotype threat on a physics test (by suggesting that no gender differences have been found on a test previously) can improve female performance. Female students performed worse in a normal physics testing situation, scoring similarly to female students in a test situation designed to elicit stereotype threat.

Although studies mainly point to the negative impact of gender stereotypes in physics and other STEM subjects, some studies have found conflicting results. \citet{Abraham_2014} found that female year 11 high school physics students did not endorse gender stereotypes, and a study by \citet{Lauer_2013} found no stereotype threat effects for female university students in introductory biology, chemistry or physics classes. The analysis employed by \citet{Lauer_2013} to assess gender differences in learning gains can be criticized for using Hakes gain \citep{Hake_1998} (see \citet{Rodriguez_2012}). That being said, their finding that all students in their sample (male and female, and in different science fields) tended to reject the stereotype that men do better than women in science is important to note. As discussed, context is important in influencing the formation of a student's habitus, and this may help explain why we see gender stereotype effects for some situations, but not others. The results of \citet{Abraham_2014} and \citet{Lauer_2013} likely suggest that the negative impact of stereotype threat occurs before students make the voluntary choice to pursue physics or other STEM subjects. Women who endorse negative gender stereotypes in physics may lack the motivation to acquire more science capital, as they believe that they cannot, or should not, study physics. 

The negative impact of gender stereotypes may be buffered by the environmental context. The perceived gender balance of a domain can influence the extent to which women identify with it. For example, \citet{Murphy_2007} found that women tended to have higher levels of belonging in STEM after witnessing a gender-balanced conference environment, rather than a male dominated one. Single-sex schools may also be a protective factor for women in physics \citep{InstituteofPhysics_2012}, and buffer against stereotype threat in mathematics \citep{Picho_2012}. 

The roles that society typically assigns to different genders is also likely to contribute towards the gender disparities in STEM. Female middle-school students tend to enjoy the social aspect of science education more than male students \citep{Dare_2016}, but physics is not often viewed as a domain that offers opportunities for social interaction or helping others. \citet{Sax_2016} found that women in physics were less likely to have a social activist orientation compared to women in other fields. Another study by \citet{Smith_2015} found that women who perceived high levels of stereotype threat in science were less likely to see science as a field that can help others. Unlike the physical sciences, areas such as health and medicine may be more likely to value the cultural capital offered by individuals who are motivated to help others. This may explain why women tend to be equally represented in life sciences. 


An individual's disposition towards science may also be impacted on by their family's view of science \citep{Rozek_2015}. \citet{Lyons_2006} found that family support was an important factor in students' choice to study physics at post-compulsory level, while \citet{Hazari_2007} found that a father's encouragement is linked to improved university physics grade for female students. Parents who value science might be more likely to talk about science at home \citep{Rozek_2015}. Being raised in a family environment where science is valued will influence the habitus of a student so that they may hold a positive disposition towards science. As suggested by \citet[p.890]{Archer_2012}, the ``combination of family habitus and capital provides a ``fertile ground'' that renders science more thinkable/desirable for their [the family's] children''. This view is supported by \citet{Hyde_2016}, who found that a mother's communication about the utility value of science subjects positively predicted an adolescent's interest in those subjects. Despite \citet{Hyde_2016} noting a lack of gender differences in the manner of mother's communication in a hypothetical scenario, the family still offers a potential source for the gender disparities observed in STEM fields.   

Factors associated with future career prospects may also influence a students habitus. Gender differences in the acquisition of science capital are pervasive, regardless of how advanced individuals are in their education. Evidence suggests that women in STEM find it more difficult than men to acquire additional science capital. This is manifested in the fact that women in STEM are likely to receive fewer resources and job opportunities than men \citep{Ivie_2013,Moss_2012}. According to \citet{Archer_2015}, the symbolic knowledge about the transferability of science in the labor market is an important form of science-related cultural capital. Female students in earlier stages of STEM education may interpret the low number of female role models employed in the STEM labor market \citep{Bray_2011} as a signal that their career prospects in these fields are limited. This could explain why \citet{Griffith_2010} found a positive relationship between the number of female STEM graduate students at a university and the level of persistence of female undergraduate STEM students, and this is likely related to an institution's commitment to ensuring an inclusive environment. 

The lack of female role models in physics may be a factor affecting the number of female physics students. Despite this, \citet{Bettinger_2005} found that having a female instructor had no significant impact on female students' interest in physics, while \citet{Hazari_2010} found that having female scientist guest speakers and discussing female scientists had no significant influence on female students' physics identity. \citet{Potvin_2016} found that the stronger a student's physics identity, the more likely they were to rate female high school physics teachers lower than male teachers. This may suggest that physics students legitimize a male embodiment of science capital over a female one.
\subsection{Confidence} 
\label{sec:5}

The previous section outlined how the gender filter is a consistent presence across the contexts of society, the family and the workplace. Each factor, from the negative gender stereotypes to the lack of female role models in physics, is likely to play a role in influencing a students habitus so they are more likely to believe that a female student can not do well in physics. This may explain why many studies find that female students tend to have lower levels of self-efficacy, self-concept and confidence in physics. 

Self-efficacy is an individual's conviction that they are able to perform a specific behavior and obtain desired results \citep{Bandura_1977}. It is a major factor related to student achievement and retention in STEM \citep{Heilbronner_2009}. Studies from the United States and Canada have found that female students tend to have lower levels of self-efficacy in STEM (including physics) during secondary school and higher education compared to male students \citep{Heilbronner_2012, Heilbronner_2009, Simon_2015}. Conversely, \citet{Louis_2011} analyzed TIMSS data from the United States and found that male high school students tended to have higher levels of self-efficacy than female students in mathematics, but not in science. \citet{MacPhee_2013} found that, although female students studying STEM subjects (including physics) at an American university tended to have lower levels of academic self-efficacy than male students (despite similar academic performance), any significant differences disappeared by graduation. 

Whereas self-efficacy is an individual's level of belief about how well they will do a specific situation, self-concept of ability is a broad measure of a student's belief in their ability in a given domain \citep{Markus_1987}, and their value of that ability \citep{Eccles_2002}. Girls who are self-confident may be more likely to intend to study STEM in the future \citep{Shapiro_2015}. However, research shows that female students tend to have lower levels of confidence and self-concept in physics\citep{B_E_2013,Sharma_2011,Hofer_2016} and mathematics\citep{Else_Quest_2013,Sheldrake_2015} compared to male students, even when there are no gender differences in grades \citep{Sheldrake_2015}. Similarly, \citet{Hardy_2013} found that boys tended to have higher levels of academic self-concept in science, with the exception of biology. 

This section has outlined the gender disparities commonly observed in STEM education, and especially physics. A brief glimpse of the current literature surrounding this topic provides an indication of it's complexity. A student's future study intentions are likely influenced from an early age, and through many different contexts. As such, a simplistic metaphor such as the leaky pipeline is not sufficient to explain a student will choose not to pursue physics. Instead, the current study makes use of a science capital framework \citep{Archer_2015}, adapted from the work of Bourdieu \citep{Bourdieu_1986}, to provide comprehensive account for why gender disparities emerge in post-compulsory education. The following sections will outline a quantitative study of physics students from the University of Auckland that used science capital as an overall research framework.
\section{The Current Study}
\label{sec:6}

The current study was motivated by the need to understand any potential gender differences in student outcomes in general, and at the University of Auckland in particular. Our study seeks to not only understand the outcomes for physics students at the University of Auckland, but to place our results in a wider socio-cultural context. We are able to achieve this goal by utilizing the work of Bourdieu\citep{Bourdieu_1985} and the science capital framework \citep{Archer_2014}. This framework enables us to use any findings to direct future research and gender equity interventions in physics.

Our analysis included all undergraduate physics courses offered by the University of Auckland. The structure of the typical physics bachelor of science degree is outlined in Table \ref{tab:Course_Structure}. Physics degrees take place over the course of three years. Students are required to pass the required courses their first year before advancing to the next year of courses. In their first year, physics students are required to take Advancing Physics 1 (AP1) and then Advancing Physics 2 (AP2) before moving onto second year. Students may also choose to take Basic Concepts for Physics in first year. This course is targeted at students who believe that they have little prior knowledge of physics, and is designed to prepare students for taking AP1 or PLS in the future. Much of our analysis focuses on the AP1 sub-population, as this is first prerequisite for more advanced physics courses; it was assumed that most students who took AP1 intended to pursue physics. 

Life science students (those majoring in biology or medicine) are required to take Physics for Life Sciences (PLS) in their first year. This means that, despite our sample including only students who took a physics course, many of the students were majoring in life science subjects. Our sample therefore allows us to compare the outcomes for students in the physics and life sciences fields. Although PLS can be taken as a substitute for AP1, it is not recommended. One significant difference between the two courses is that AP1 assumes a knowledge of calculus, while PLS does not. The current study was able to compare the AP1 and PLS subsets of the general physics population to account for a student's first year STEM intentions.

\subsection{Aims}
\label{sec:7}

The current study used quantitative analysis to investigate gender differences in University of Auckland physics outcomes. Our results are then interpreted using the science capital framework; this allows us to understand any gender differences in a wider socio-cultural context. Based on previous literature, three specific research questions were investigated.

\subsubsection{Subject enrollment.}
\label{sec:8}

Are there gender differences in subject enrollment after first year physics? Based on literature outlined previously \citep{EducationCounts_2016b,Kennedy_2014}, it was expected that female students would be more likely to take life science subjects (biology and medicine), while male students would be more likely to take the physical science subjects (physics, mathematics, engineering and computer science). 

\subsubsection{Physics Performance.} 
\label{sec:9}

Are there gender differences in physics course performance? Based on previous literature \citep{Ost_2010,Smith_2011}, it was expected that gender differences in course performance would be minimal after controlling for high school performance and the type of high school that a student attended. Previous research suggests that single-sex schools may improve performance for female physics students \citep{InstituteofPhysics_2012,Picho_2012}.

\subsubsection{Confidence.} 
\label{sec:10}

Are there gender differences in the enrollment patterns of students entering an introductory physics course (Basic Concepts for Physics), before taking a prerequisite course for a physics major (AP1)? Basic Concepts of Physics is designed for students who believe they little prior knowledge of physics, to prepare them for more advanced physics courses \citep{UoA_2016b}. We make the assumption that, after controlling for academic preparation, students who are more confident in their physics ability will skip Basic Concepts of Physics before taking AP1, while less confident students will be more likely to take Basic Concepts of Physics before taking AP1. Alternatively, academically unprepared students who did not take Basic Concepts for Physics might be over-confident in their ability. Based on previous research regarding self-efficacy and confidence \citep{Heilbronner_2012,MacPhee_2013}, it was expected that female AP1 students, regardless of academic preparation, would be more likely to take Basic Concepts for Physics beforehand.

\section{Methodology}
\label{sec:11}

\subsection{Dataset}
\label{sec:12}

The current study utilized student records for 10,874 unique students who had enrolled in at least one physics course at the University of Auckland between semester one, 2009 and semester one, 2015. Of these students, 10,764 had demographic (e.g., gender, high school attended) and academic (subject, grade, etc.) information. Data included records of non-physics courses, as long as a student had enrolled in at least one physics course during the study period. Student records from semester one, 2015 were not included in analysis due to them not being complete. Students with incomplete records were also excluded from analysis, leaving 9,954 unique students. 

Overall, 40\% of the 9,954 students in our dataset were female. This percentage drops to around 30\% when PLS (Physics for Life Sciences) students are excluded. 7868 students went to co-educational schools (34\% female), while 2086 students went to single-sex schools (63\% female). Female students tended to enter first year physics with higher GPE scores (Figure \ref{fig:GPE}).
\subsection{Measures}
\label{sec:13}

The following variables were used in the analysis:
\paragraph{Grade Point Equivalence (GPE).} GPE is an entry level score that provides a standard measure of a student’s prior academic performance at the time of admission, regardless of the qualification they previously took. It is measured on a 0-9 scale, with 9 being the highest performing. It provides an aggregate measure of how well a student did in all of their high school courses \citep{UoA_2016a}.

\paragraph{Grade Point Unit (GPU).} GPU is a measure of a student’s university performance in a single course. It is measured on a 0-9 scale, with 0 being equivalent to a fail (D+ or lower), and 9 being equivalent to an A+ grade. 
	
\paragraph{Gender.} Due to data limitations, gender was only recorded as male or female. Female students were assigned a value of 1, and male students were assigned a value of 0. 
	
\paragraph{High School Type.}  High school type indicates whether or not the student went to a single sex or a co-educational high school prior to university. High school type was recoded into binary values, with co-educational schools being assigned a value of 1, and single-sex schools a 0. 

\subsection{Procedures}
\label{sec:15}

The current study employed different quantitative analytical procedures for each research question.

\subsubsection{Subject enrollment} Are there gender differences in subject enrollment after first year physics? Chi-square tests were used to investigate the role of gender in subject selection at second year and third year. The domains analyzed were physics, mathematics, engineering, computer science, chemistry, biology and medicine. Odds ratios with confidence intervals were calculated to determine the likelihood of a female student being enrolled in a subject at second year and third year compared to a male student, after taking physics at first year. Confidence intervals were generated using the formula
\begin{equation*}
\begin{split}
\mbox{(Lower Bound, Upper Bound)} = \\ \mbox{LN(OR)} \pm z \times \mbox{SE LN(OR)}
\end{split}
\end{equation*}
where z corresponds the degree of confidence (1.96 for 95\%, 2.576 for 99\% etc.). 
This procedure was first conducted using all physics students, and then repeated for two subpopulations; those taking AP1 and those taking PLS. 

\subsubsection{Physics Performance} Are there gender differences in physics course performance, after controlling for potential confounding variables? In order to assess the differences in course outcomes for male and female physics students, a linear regression model with GPE predicting course GPU was used. To assess whether this relationship was moderated by gender, a GPE$\times$gender interaction term was added to the model. This process was repeated for all first year and second year physics courses. The AP1 subpopulation was analyzed in more depth, adding high~school~type interactions to the original model. Gender differences in physics course performance may explain gender differences in subsequent subject enrollment. Alternatively, course performance may differentially influence the study paths of male and female students, even after accounting for grade.

\subsubsection{Confidence} Are there differences in the enrollment patterns of students entering an introductory physics courses (Basic Concepts of Physics) before taking a prerequisite course for a physics major (AP1)? The number of academically prepared and academically unprepared students who took introductory courses before taking AP1 were compared. An academically prepared student was defined as one who had a GPE above three, which is equivalent to a C+, and a academically unprepared student as one who had a GPE equal to or below a three. This threshold was used as many courses require previous grades that are above a C+. After splitting students into academically prepared and unprepared blocks, $2\times 2$ contingency tables were formulated, comparing gender (male or female) with whether or not they took Basic Concepts for Physics. Odds ratios were then calculated with confidence intervals of 95\%.

\section{Results}
\label{sec:16}

\subsection{Subject enrollment}
\label{sec:17}

Of the students who enrolled in a physics course in their first year, male students were generally more likely to study the physical science subjects in later years, while female students were more likely to study life science subjects. The odds of a female first year physics student taking a subject in later years over a male student are reported in Table \ref{tab:RQ2}. Odds (and their associated standard errors) are reported for the overall physics population, and the AP1 and PLS sub-populations. Figure \ref{fig:S2_OddsRatios} highlights the odds of a female student enrolling in a subject in second year over a male student. 

Female first year physics students were around 2.6 times more likely to take biological science (biosci), 2.5 times more likely to take medical science (medsci) and 1.9 times more likely to take chemistry in their second year than male students. Results show that this pattern is present for the AP1 students, with female students being approximately 8.8 times more likely to enroll in medsci at second year than male students. Of the PLS students, female students were still more likely to enroll in the life science subjects in later years than male students (with the exception of chemistry at third year).  

Table \ref{tab:RQ2} shows that for the AP1 students, there tended to be no significant differences in second year and third year physical sciences enrollments. However, male students were consistently more likely to enroll in computer science in their second year and third year, regardless of subpopulation. Overall, male students were around 5 times more likely to take computer science in their second year compared to female students. Male AP1 students were around 2 times as likely to take computer science in their second year compared to female AP1 students. On the other hand, male PLS students were around 1.8 and 2 times more likely to take mathematics and physics, respectively, in their second year. Male PLS students were also more likely to enroll in computer science and engineering courses in their second year and third year than female students.

\subsection{Physics Performance}
\label{sec:18}

Overall means and standard deviations are reported in Table \ref{fig:Means_SD}. A multilevel model was developed to predict first and second year physics course grades based on GPE and gender (Table \ref{tab:RQ2_1}). The multilevel model allowed us to control for the hierarchical nature of our data, where student course grades are nested within a specific physics course, and time of study. The results show that, overall, GPE significantly and positively predicted physics course grades ($\beta$ $=$ 0.502, SD $=$ 0.017, p $<$0.001). Gender differences were minimal and insignificant in our final model ($\beta$ $=$ 0.075, SD $=$ 0.122, p $=$ 0.542), although gender differences in grades varied considerably between courses (random effects SD = 0.461). Figure \ref{fig:MLM_LinearModels} visualizes linear models where course grade is predicted by GPE as a function of gender. We are able to see that gender differences in course grades were more pronounced in first year courses, with female students with low GPEs tending to underperform compared to male students with low GPEs. However, in second year courses this trend is reduced, and if anything reversed. 

In depth analysis focused on the subpopulation of students who enrolled in AP1, as this course is a prerequisite for advanced physics courses. A multiple linear regression model showed that GPE and gender significantly predicted AP1 grade, while an interaction between both factors tended towards significance (Table \ref{tab:RQ2_1}). 

The fitted regression model was:
\begin{equation*}
\begin{split}
\mbox{AP1 GPU} = 0.12 +  0.82\mbox{ GPE}\\-0.48\mbox{ gender} + 0.07\mbox{ GPE}\times \mbox{gender} 
\end{split}
\end{equation*}

where GPE was measured on a 0--9 scale and gender was coded as 0 = male and 1 = female. AP1 grade increased 0.82 units for every 1 increase in GPE, decreased 0.48 units for female students, and increased 0.07 units for every 1 unit increase of GPE$\times$ gender. In more basic terms, as a student's GPE increased, AP1 grade was likely to increase. On the other hand, female students tended to have lower AP1 grades than male students, with the exception of students with high GPE levels.  

A second model was calculated to predict AP1 grade based on GPE, gender and high school type. In this model, GPE, gender, high school type and the GPE$\times$ high school type interaction were significant predictors of AP1 grade (Table \ref{tab:RQ2_LM}). The fitted regression model was:

\begin{equation*}
  \begin{split}\mbox{AP1 GPU} = 2.957 +1.890\mbox{ GPE (centered)} \\-0.307\mbox{ gender}\\ + 0.411\mbox{ high school type} \\+ 0.035\mbox{ GPE (centered)}\times\mbox{gender}\\  -0.535\mbox{ GPE (centered)}\times\mbox{high school type}\\  -0.187\mbox{ gender}\times\mbox{high school type} \\+  0.135\mbox{GPE (centered)}\times\mbox{gender}\times\mbox{high school type} 
  \end{split}
\end{equation*}
where high school type was coded as 0 = single-sex and 1 = co-educational.

As seen in Figure \ref{fig:A}, of the students with lower GPE scores, male students, and students from co-educational schools, tended to get higher AP1 grades. Gender differences were minimal for students with higher GPE scores. Instead, we see that, among students with high GPE scores, those who attended single-sex schools tended to get higher AP1 grades than those from co-educational schools.

\subsection{Confidence}
\label{sec:19}

To see whether there were gender differences in the number of academically prepared students who took Basic Concepts of Physics before AP1, $2\times 2$ contingency tables were made (Table \ref{tab:RQ3}). This procedure was repeated for academically unprepared students (Table \ref{tab:RQ3_2}). 

Few academically prepared students (GPE $>$ 3) took Basic Concepts of Physics before taking AP1, but sample size was sufficient to conduct a chi-square test. The relationship between gender and taking Basic Concepts of Physics was significant, $ \chi 2 (1, N = 810) = 7.03, p = .008$. Female students were around 4.98 times more likely to take Basic Concepts of Physics before taking AP1 ({Table \ref{tab:RQ3_3}).

Of the academically unprepared students, a chi square test indicated that the relationship between gender and taking Basic Concepts of Physics tended towards significance, $ \chi 2 (1, N = 609) = 3.69, p = .054$. Females students were around 1.74 times more likely to take Basic Concepts of Physics before AP1 ({Table \ref{tab:RQ3_3}). Despite the previous tests not being a direct measure of confidence, these results are consistent with the idea that female students tend to be less confident in physics.

\section{Discussion} 
\label{sec:20}

\subsection{Subject enrollment}
\label{sec:21}

This study found gender differences in subject enrollment, proving the hypothesis that male first year physics students would be more likely to enroll in physics, mathematics, and computer science courses in later years, while female first year physics students would be more likely to enroll in biology, medicine science and chemistry. This is consistent with international research \citep{Kennedy_2014,Smith_2011,Stevanovic_2013}. 

The influence of factors located in the socio-cultural contexts of each student is likely to be responsible for the differences in representation seen in the current study. Each student's habitus is influenced by what the student experiences in their everyday life. Due to the influence of societal (i.e., gender roles, stereotypes), family, school and employment factors (i.e., lack of resources and role models), a student's habitus is likely to be influenced from an early age to be congruent with what is stereotypical for their gender. The student's habitus then generates the student's practices in the field of study, and the student's appreciation of that field. Although we do not claim that every student has the same habitus, there is clear evidence that stereotypes regarding women and science pervade society \citep{Nosek_2009}. Appearing feminine may result in female students being more likely to perceive that science, and especially physics, is a masculine domain that does not fit a stereotypical feminine identity \citep{Archer_2013}. The current results suggest that life sciences are more appealing to female students, with this evidenced by the fact that the proportion of female students in physics drops from 40\% to 30\% after excluding all PLS students. Compared to other STEM domains, the life sciences may be more positively perceived by female students due to the lack of negative gender stereotypes and the increased availability of female role models, among other factors. The life sciences also offer the opportunity to work with and help others, which, according to prior research, may be more important for female students compared to male students \citep{Su_2009, Archer_2013}. On the other hand, the perceived masculine culture of physics may be incongruent with a female student's habitus.

Unfortunately, this study did not have access to information regarding high school subject selection to investigate how high school course selection impacts on university course selection. Instead, AP1 enrollment was used as a proxy for intent to major in physics, and PLS enrollment as a proxy for intent to major in a life science degree.

The gender differences in second year physics and mathematics enrollment disappeared when taking AP1 enrollment into account. Female AP1 students are likely to have cultivated an identity in physics, and this idea is supported by research suggesting that individuals who identify with physics are more likely to progress in the domain \citep{Hazari_2010}. Female students who have high levels of physics identity tend to have lower social-activist orientations and less desire for family time than women in other fields \citep{Sax_2016,Hazari_2010}. It may be that progression in physics requires an assimilation to a masculine culture that conflicts with desires traditionally held by women \citep{Gonsalves_2016}, such as the desire to help others \citep{Su_2009}. Hence, the masculine culture of physics is likely to disproportionately filter out female students from physics. Future studies can explore this argument further by going beyond a binary interpretation of gender to instead look at, for example, gender identity.

The results in Table \ref{tab:RQ2} show that female AP1 students were more likely to take chemistry, biology, and medicine in their second year compared to male AP1 students. The gender differences seen for AP1 students in chemistry, medicine and computer science were maintained in their third year. This suggests that, of the students who initially intended to major in physics, female students were more likely than male student to either switch out of physics and pursue life sciences, or take a physics-life sciences conjoint degree. 

The change of STEM field may be interpreted as what Bourdieu called a ``transverse movement'' \citep[p.131]{Bourdieu_1985}. A transverse movement to another field may occur when an individual wants to prevent their capital from being diminished. A female AP1 student's initial intentions to major in physics may have changed if they feel that their science capital may be more valued in the life sciences. Thus, a female AP1 student may be more likely to conclude that the investment of their science capital in the field of physics will not be as profitable compared to an investment in another field (i.e., life sciences). To prevent the transverse movements of female students away from physics, the culture of physics needs to ensure that every student's capital is equally valued.  
    
Even when both male and female students shared the original intentions to pursue life sciences (based on PLS enrollment), female students were more likely than male students to maintain, or fulfill, these intentions in their second year and third year. Male PLS students were instead more likely to enroll in the physical science subjects. It may be that male students find that the life sciences do not offer the same translation of science capital that they find in the physical sciences, and thus switching to physics is justifiable educational choice. The above findings suggest that the physical sciences continue to be culturally reproduced as a male domain at university level.

\subsection{Physics Performance}
\label{sec:22}

Table \ref{tab:RQ2_1} shows the results of a multilevel model used to predict physics course grades after controlling for factors relating to physics courses (i.e., the course taken and the year when the course was taken) and student academic preparedness (measured by GPE). We found that, overall, there were mostly small gender differences in physics performance, and that gender differences varied considerably by physics course. Male students tended to have higher grades on average in first year physics courses. However, in second year courses differences are minimal, and if anything, the trend seen in first year courses is reversed. These findings suggest that, to a certain extent, the gender differences in science performance that favor male students in New Zealand high schools \citep{Martin_2012,May_2013,May_2016} persist in the first year of university physics. We found that male students had higher grades on average in AP1, but not in courses that physics majors would take in the next stages of their degrees (i.e., AP2, and year two physics). It may be that female students who do poorly in AP1 drop out of the physics pipeline at a greater rate than their male counterparts, and this is what causes the mean differences between male and female students to shrink, or even reverse, in more advanced courses. 

Further analysis of AP1 showed that, of the students with low GPE scores, male students, and students from co-educational schools, tended to have slightly better AP1 grades (Figure \ref{fig:A}). There were no gender differences present for students with high GPE scores, while students with high GPE scores tended to get slightly higher AP1 grades if they were from a single-sex school rather than a co-educational school. This finding provides weak evidence that single-sex schools positively impact on university performance if a student does well at high school; others may find the transition to university physics more difficult. It could be that negative factors, such as stereotype threat \citep{Picho_2012}, emerge with the transition from a single-sex environment to a male dominated one. Individuals who did well at school may have a strong academic identity which protects them from these factors, or a habitus that positively disposes them to further physics study.

To understand why the relationship between a student's GPE and AP1 grade was moderated by school type, future studies should control for high school characteristics. Although we cannot be sure in this study, it may be the case that the school type influence we observed is actually the result of a school socio-economic status. Students from high socio-economic backgrounds are more able to access the cultural capital that is distributed by society \citep{Bourdieu_1973}. These students would have increased access to science-related resources, equipment, and economic capital which can be transformed into educational achievement. The impact of socio-economic background on student outcomes in physics is another equity issue, and needs to be controlled for when assessing the impact of school background on performance.

\subsection{Confidence}
\label{sec:23}

Basic Concepts of Physics is designed for students who believe they little prior knowledge of physics, to prepare them for more advanced physics courses \citep{UoA_2016b}. We made the assumption that, after controlling for academic preparation, students who are more confident in their physics ability will skip Basic Concepts of Physics before taking AP1, while less confident students will be more likely to take Basic Concepts of Physics before taking AP1. The current study found that academically prepared female students were more likely to take Basic Concepts of Physics before taking AP1, and this may signal that they were not confident in their ability to do well in an advanced physics course. Although not a direct measure of confidence, the above conclusion is consistent with previous research that suggests that female students tend to have lower confidence \citep{Sharma_2011} and self-concept in physics \citep{B_E_2013, Hofer_2016}.
    
Gender differences in confidence provide an explanation as to why any gender differences in second year physics course performance favored female students. Less academically prepared male students were less likely to take Basic Concepts of Physics before AP1, which may signal over-confidence in physics. If these male students remained over-confident, they may still progress to second year physics despite poor grades. On the other hand, female students may lack the confidence or self-efficacy to continue in the face of poor grades. This would change the gender-ratio of the students represented in low and high achieving groups that were originally seen in first year.

The GPE measure of high school performance is limited as it is an overall measure of high school performance, and not subject specific. Female students may be more likely to perform poorly in high school physics, or opt out of high school physics altogether. In this case, taking Basic Concepts of Physics before AP1 would be a realistic course selection choice, and it would not signal a lack of confidence. However, it is unlikely that a student would opt out of physics at high school, yet enroll in it at university. To fully understand the transition from high school to university, more detailed high school data is needed.
 
\subsection{Implications}
\label{sec:24}

The current study contributes to the evidence of gender disparities in physics education at university. Furthermore, the current study demonstrates the utility of the science capital framework as an interpretive lens. This approach carries many benefits over other, more simplistic models, such as the leaky pipeline model. Firstly, the science capital framework allows us to place our results in a broader socio-cultural context. Doing so removes any stigma that can be attached to students who `leak' from the physics education pipeline. Instead, the science capital framework places emphasis on the impact of contextual factors on student's course selection. Further to this, emphasizing the contexts that students are situated in allows us as researchers to formulate suitable interventions to boost the physics enrollments of underrepresented groups. 

The findings of the current study suggest that there are subtle gender differences in both physics performance and course selection that point towards more systematic gender disparities in physics education. These findings indicate that more needs to be done to ensure that physics is perceived as a viable option for female students. This can be done by using interventions to boost the value that science capital holds in all areas of society. Echoing the arguments of \citet{Claussen_2013}, science education in New Zealand, and internationally, needs to highlight the utility value of physics in culture, scientific literacy, and employment. We must also make sure that Based on previous research, female students may benefit from increased exposure to female role models in physics, and examples of how physics can help people. 

Most importantly, we need to ensure that physics maintains a culture that places value in the science capital offered by all groups of students. Based on our findings, interventions to boost the number of female students graduating in physics would be most useful at stages of education prior to university. It is likely that by university, student's habitus is less susceptible to change. In order to generate a more inclusive environment in physics at the University of Auckland, it may be necessary to implement an initiative similar to Athena SWAN \citep{AthenaSwan} to boost gender equity. Athena SWAN has been shown to have a positive impact in boosting gender equity in STEM in the United Kingdom, Ireland and Australia \citep{Ovseiko_2017} by encouraging institutions to adopt changes that boost the representation of women in STEM, remove obstacles that face women in STEM, and provide a better working environment for all staff\citep{AthenaSwan}.

Despite the utility of using Bourdieu's theory to place quantitative data in context, more research is needed to build on this approach. Future studies should attempt to replicate the methodology of \citet{Archer_2014}, specifically employing qualitative analysis in order to directly study students' science capital and science aspirations. It can be argued that due to differences in concepts, activities and careers related to these fields, separate forms of science capital may need to be conceptualized. Certain aspects of science capital may be important for the physical sciences, but not the life sciences and vice versa. For example, a student's appreciation of mathematics may be a key component of physical science capital, while a desire to help others may be more important for life science capital.

\subsection{Limitations}
\label{sec:25}

There were several limitations to the current study. Our dataset was limited to university course selection and performance information, and student demographic information. We did not have data regarding the course selection information of students prior to university. Such data would have provided more detail about  students' educational trajectories. Our data was also limited by the fact that all students had to have enrolled in at least one physics course to be included in our dataset.  In future work, we plan to extend the current analysis across other departments, including those from the life sciences.

Our study would have also benefited from combining our quantitative analysis with qualitative measures. Surveys and interviews of students would provide contextual and fine-grained detail that would compliment our quantitative analysis. Using the above procedures would also enable us to obtain a non-binary measure of a student's gender identity and to assess what role this may play in course selection and performance.

Despite having access to information regarding the ethnicity of students, we decided against using this information in our analysis. This is due to the fact that preliminary analysis showed extremely low cell sizes for ethnic groups other than New Zealand European and Asian students, in particular M\={a}ori and Pasifika students. When possible, future studies should make use of an intersectional research design (one that explores the interaction between gender, ethnicity, social class etc.). This is especially important when using a Bourdieusian framework to interpret results. As suggested by Bourdieu: ``The individuals grouped in a class that is constructed in a particular respect... always bring with them secondary properties'' \citep[p.102]{Bourdieu_1985}. Understanding the intersection of student characteristics would allow us to control for the secondary properties that Bourdieu speaks of. To understand why there were low cell sizes for minority groups, future studies should seek to analyze data from earlier educational stages (e.g., the transition between compulsory and post-compulsory science education at high school).

\section{Conclusion}
\label{sec:26}

The current study investigated gender differences for University of Auckland physics students by using a quantitative analysis of student data. Results showed gender differences in subject enrollment consistent with gender stereotypes: after taking a first year physics course, female students were around 2.5 times more likely to study life science subjects (medicine and biology) in later years, while male students were around 2 times more likely to study physical sciences, compared with female students. With regards to physics performance,gender differences were small overall, but varied considerably between courses. Male students tended to get slightly higher grades than female students in first year, but this trend was not present in second year and if anything, was reversed. At low levels of academic preparedness, male students and those from co-educational school outperformed those from other groups in AP1. Of students with a higher level of academic preparedness, those from single-sex schools outperformed those from co-educational schools, while gender was not a significant predictor of AP1 grade. It is likely that this finding is linked to the availability of science capital for students attending different schools.

Academically well prepared female students were around 5 times more likely to take an introductory course (Basic Concepts for Physics) before a progressing physics course (AP1), compared to equivalent male students. This suggests that female students may be more likely to lack confidence in physics, and that students' self-concept in physics may be moderated by gender. Through the lens of science capital, the combination of results offered by the current study suggest that the life sciences may place more value on the science capital offered by female students. It is likely that a student's habitus in relation to the field of physics disproportionately discourages female students from aspiring to study in this field. In turn, this may feed into the cultural reproduction of physics as a male domain. More needs to be done to ensure that all types of science capital are equally valued in society, the family and in future physics careers. Boosting the enrollment rates of female students making the transition from high school to progressing physics courses at university level (such as AP1) should improve the perceived value of science capital for female students.

\clearpage

\section{}

\begin{figure*}
\begin{center}
\begin{tikzpicture}[scale=1.045]

\node at (1.5,-0.0) {\textbf{Science Capital}};
\node at (3.3,-0.0) {\textbf{$\times$}};

\node at (4.5, -0.0) {\textbf{\textbf{Habitus}}};
\draw[very thick, ->] (5.75,-0.0) -- (8.0,-0.0);
\node at (6.0, 2.5) {\textbf{\textbf{Field of Study}}};

\node at (10,0.5) {\textbf{Science Outcomes}};
\node at (10,-0.0) {Grade Point Unit};
\node at (10,-0.5) {Course enrollment};

\draw[black, very thick] (6,0) ellipse (8 and 3);
\end{tikzpicture}
\end{center}
\caption{\label{fig:Model} The science capital frame work used in the current study, adapted from the original model outlined by \citet[p.10]{Bourdieu_1985} and the work \citet{Archer_2015}. A student's habitus interacts with their acquired level of science capital to produce an individual's science outcomes. Student's habitus is formed in relation to the specific socio-cultural context of a field, and generates both the students practices in the field, and their appreciation of the field. An individual who is positively predisposed to study in a scientific field, whilst also having access to various forms of science capital, will likely achieve higher grades in that field and aspire to study that in that field in the future.}
\end{figure*}
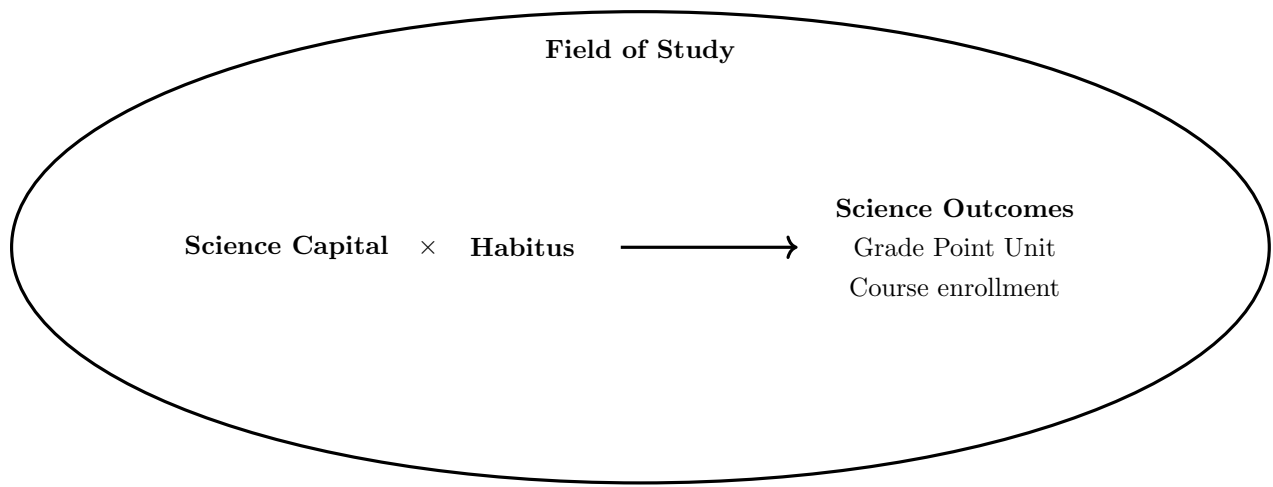

\begin{turnpage}
\begin{table*}
\begin{center}
\begin{tabular}{ |c|c|c| } 
 \hline
 First Year &  Second Year &  Third Year \\ 
  \hline
 Basic Concepts for Physics & Mathematical Methods for Physics & Classical and Statistical Physics \\ 
\textbf{ Advancing Physics 1} & The Geophysical Environment & Electromagnetism \\ 
\textbf{ Advancing Physics 2} & \textbf{Electromagnetism and Thermal Physics} & Optics and Laser Physics \\
 \textbf{Physics for Life Sciences} & Materials and Waves & Electronics and Signal Processing \\
 Introductory Physics for Science and Engineering & \textbf{Classical Physics} & Quantum Mechanics and Atomic Physics \\
 Planets, Stars and Galaxies & Networks and Electronics & Condensed and Soft Matter Physics \\
 Science and Technology of Sustainable Energy & \textbf{Quantum Physics} & Particle Physics and Astrophysics \\
 Analytical Techniques in Physical Sciences & \textbf{Modern Physics} & \textbf{Experimental Physics 1} \\
 Properties of Matter & Optics &\textbf{ Experimental Physics 2 }\\
 Digital Fundamentals & Optics and Electromagnetism &  \\
  & Medical Physics &  \\

 \hline
\end{tabular}
\end{center}
\caption{\label{tab:Course_Structure} The above table describes the Physics courses offered by the University of Auckland, with key courses in bold. Physics majors must pass Advancing Physics 1 and Advancing Physics 2 in first year before they can progress to second year. They must then pass at least 4 courses at second year before they can take the more advanced third year courses. To be eligible for graduation, students must then pass 4 courses at third year level, with Experimental Physics 1 or 2 as a prerequisite. Physics for Life Sciences is a prerequisite first year course for students majoring in a life science subject (medicine or biology).}
\end{table*}
\end{turnpage}

\begin{figure*}[hp]
\centering
\includegraphics[width=15cm, height = 10cm]{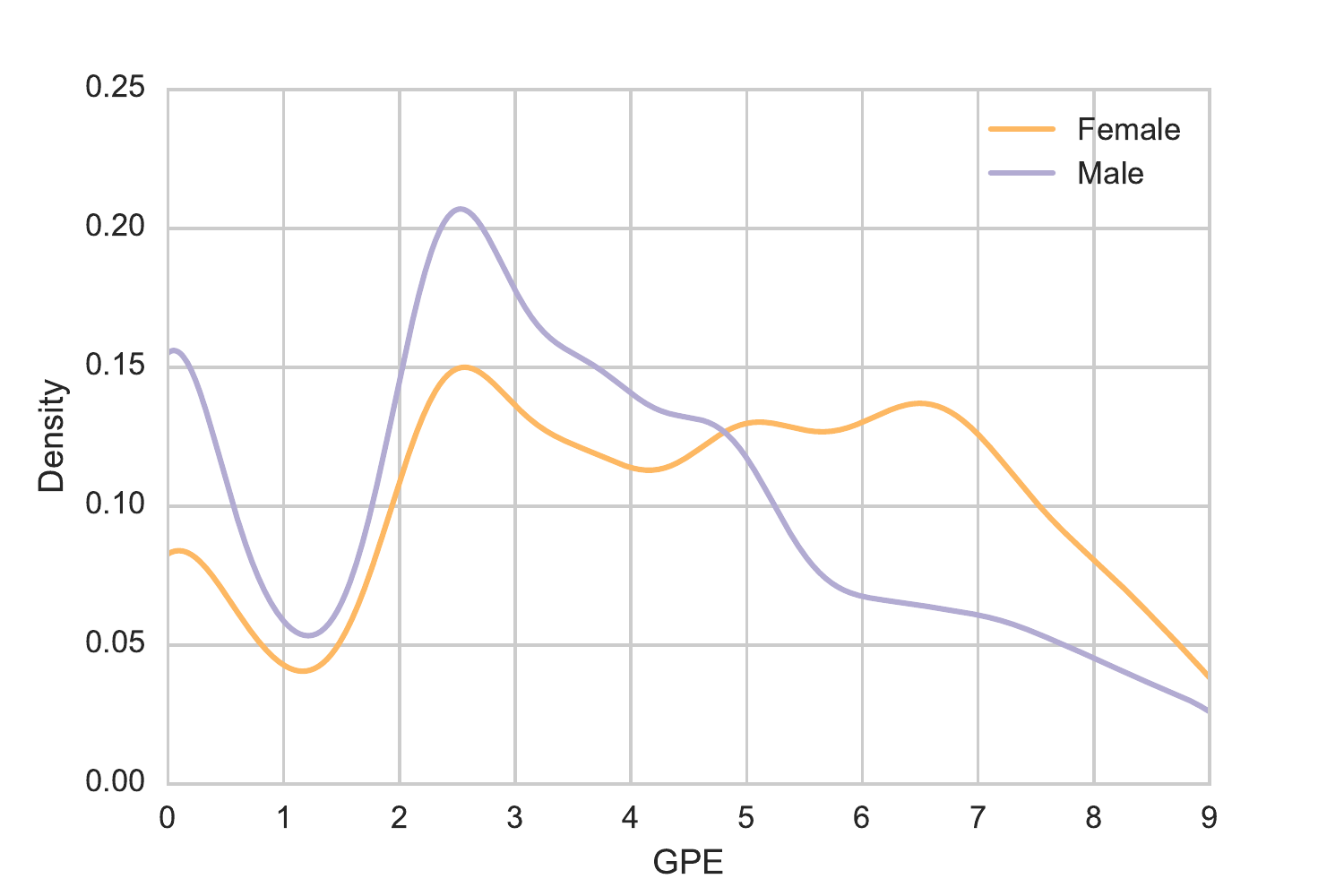}
\caption{\label{fig:GPE}A density plot showing the GPE distribution of male and female physics students. For both male and female students, GPEs tend to peak around 2.5. Female students tended to have a higher density of GPEs located in the upper range of 6-9 compared to male students}
\end{figure*}

\begin{turnpage}\centering
\begin{table*}
  \begin{tabular}{@{\extracolsep{5pt}}cclllllll}\\ \toprule
  & & \multicolumn{7}{ c }{Subject} \\ 
  & & \multicolumn{1}{c}{Compsci} & \multicolumn{1}{c}{Physics} & \multicolumn{1}{c}{Engineering} & \multicolumn{1}{c}{Maths} & \multicolumn{1}{c}{Chem} & \multicolumn{1}{c}{Medsci} & \multicolumn{1}{c}{Biosci}\\ \cline{1-9}
  \multicolumn{1}{ c  }{\multirow{4}{*}[2ex]{Second Year} } &
  \multicolumn{1}{ c }{All Students} & 0.23 (0.08)*** & 0.40 (0.09)*** & 0.43 (0.12)*** & 0.44 (0.06)*** & 1.90 (0.08)*** & 2.52(0.06)*** & 2.62 (0.06)*** \\ 

  \multicolumn{1}{ c  }{}                        &
  \multicolumn{1}{ c }{AP1} 	& 0.48 (0.15)***	&	0.87 (0.13)	&	0.89 (0.20)&	0.97 (0.12) &		2.03 (0.21)*	&	8.76 (0.54)***	&	2.30 (0.35)*	  \\
  \multicolumn{1}{ c  }{}                        &
  \multicolumn{1}{ c }{PLS} & 0.30 (0.16)***	&	0.51 (0.18)***	&	0.33 (0.20)*** &	0.56 (0.10)*** &		1.48 (0.10)***	&	1.47 (0.07)***	&	1.54 (0.07)***	 \\ \cline{1-9}

\multicolumn{1}{ c  }{\multirow{4}{*}[2ex]{Third Year} } &
\multicolumn{1}{ c }{All Students} & 0.22	(0.10)***	&	0.55 (0.14)***	&	0.44 (0.15)*** &	0.52 (0.11)*** &		1.89 (0.09)***	&	2.30 (0.07)***	&	2.57	(0.07)***	
\\ 
\multicolumn{1}{ c  }{}                        &
                      
\multicolumn{1}{ c }{AP1} & 0.39	(0.20)*** &	1.03 (0.19)  &	0.85	(0.22) &	1.07 (0.18) &	2.65 (0.27)** &	7.18 (0.71)**	&	1.52 (0.49)   \\ 
\multicolumn{1}{ c  }{}                        &
\multicolumn{1}{ c }{PLS} & 0.37 (0.21)*** &	0.84 (0.28) &	0.35	(0.25)*** &	0.58 (0.26)* &	1.20 (0.11)	&	1.35 (0.07)*** &	1.48 (0.07)***  \\ \botrule

\end{tabular} 

\caption{\label{tab:RQ2} The above table presents the odds (and SE of log odds) of a female student taking a subject after first year over a male student. Odds ratios above one signify that a female student was more likely to take that subject, while odds ratios below one signify that a male was more likely to take that subject. Columns are roughly ordered from most male subject to most female subject based on the odds ratio for all students. The proportions of female students comprising each population are as follows: All Students: 40\%, AP1 Students: 22\%, PLS Students: 54\%. $^{*}$p$<$ .05; $^{**}$p$<$.01; $^{***}$p$<$.001. Second year subject enrollments are summed up in \ref{fig:S2_OddsRatios}}.

\end{table*}
\end{turnpage}

\begin{table*}[hp]
\centering
\begin{tabular}{lcccc}
\hline\hline
Physics Course & N & Mean Grade & SD  \\ \hline
Basic Concepts & 1508 & 3.117 & 2.963  \\
 Intro to Astro &	1248&	4.377	&3.018  \\
Intro to Astro (GenEd)	&1007	&4.096	&3.007 \\
	Advancing Physics 1	&1964&	3.266&	2.947\\
	Properties of Matter	&695&	3.658&	3.118\\
	Digital Fundamentals	&801	&4.426&	3.160\\
	Advancing Physics 2	&1153&	3.232&	3.020\\
	Physics for Life Sciences&	4672&	5.085&	3.075\\\hline
	Mathematical Methods	&312&	2.372	&2.921\\
	Geophysics	&231&	4.126&	2.863\\
	Electromagnetism/Thermal&	106&	2.217&	3.080\\
	Materials/Waves&	98&	3.296&	3.037\\
	Classical Physics&	526&	2.947&	3.158\\
	Electronics	&407&	3.032&	3.120\\
	Quantum Physics&	81&	3.642&	3.411\\
	Modern Physics&	295	&3.668	&3.309\\
	Optics&	51&	2.686	&2.709\\
	Optics/Electromagnetism	&305&	3.141&	3.124\\
	Astrophysics&	38&	3.526	&3.244\\
	Medical Physics&	88&	4.489&	2.869\\\hline\hline
\end{tabular}
\caption{\label{fig:Means_SD} A table showing the number of students in each course between 2009 and 2015, and mean grades and standard deviations. The first section of the table shows the courses offered in first year, and the second section shows the courses offered in the second year. The table shows that the average grades in most courses ranged between 3 (letter grade equivalent being a C+) and 5 (letter grade equivalent being a B), with Mathematical Methods having the lowest average grade with 2.372 and Physics for Life Sciences having the highest average grade with 5.085. The table also shows that the number of students enrolled in second year physics courses is a lot lower than the numbers seen in first year courses.}
\end{table*}

\begin{figure*}[hp]
\centering
\includegraphics[width=15cm, height = 15cm]{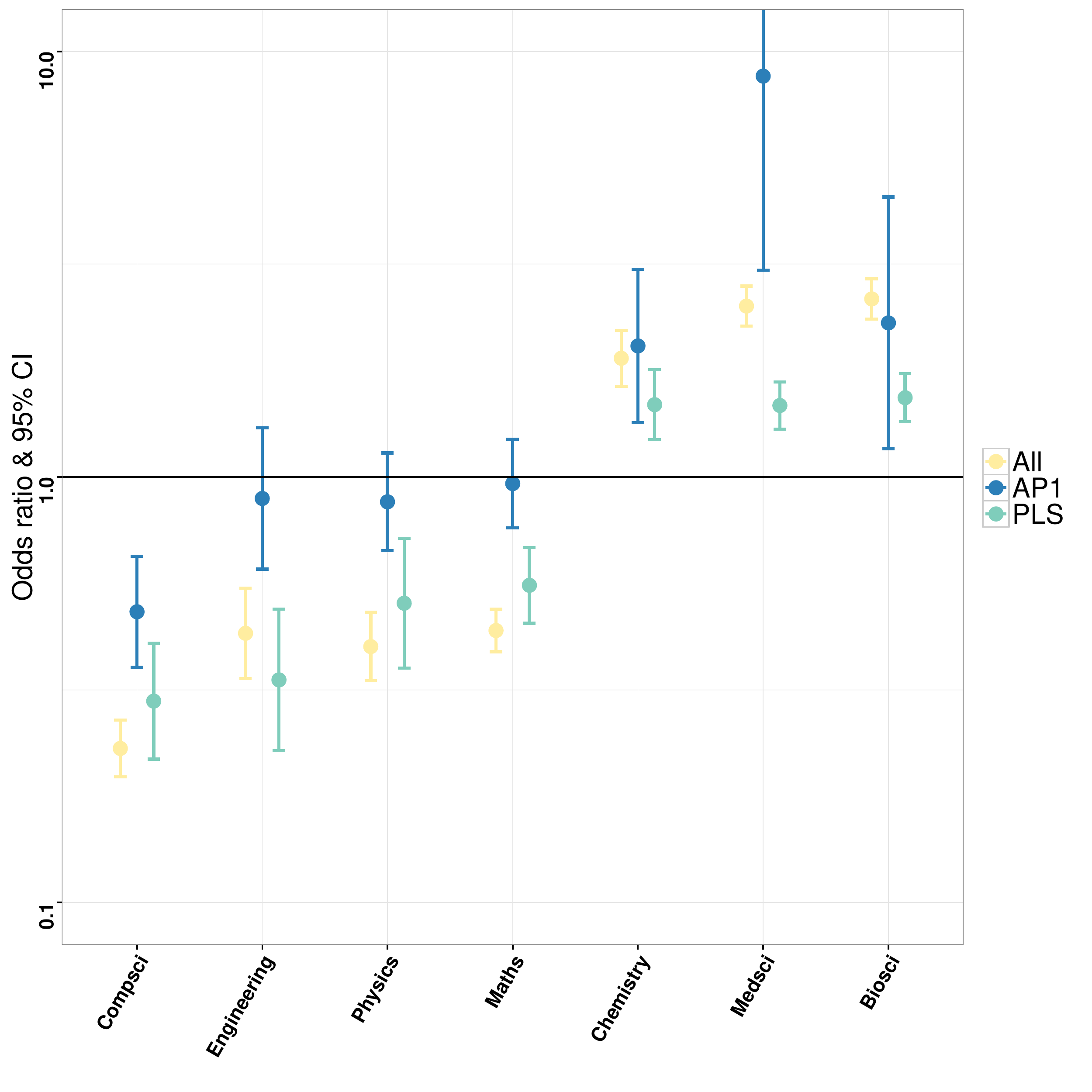}
\caption{\label{fig:S2_OddsRatios}A plot showing the odds ratios (and associated 95\% confidence intervals) of physics students subject selection in their second year. Odds ratios are presented for all students, AP1 students and PLS students. This figure shows a clear trend where female students tend to be more likely to take life science subjects in their second year compared to male students. Male students tend to be more likely to take computer science in their second year. With the exception of students from the AP1 subpopulation, male students tend to be more likely to take engineering, physics and maths in their second year compared to female students.}
\end{figure*}

\begin{turnpage}
\begin{table*}
\begin{tabular}{@{\extracolsep{5pt}}lD{.}{.}{-3} D{.}{.}{-3} D{.}{.}{-3} D{.}{.}{-3} D{.}{.}{-3} D{.}{.}{-3} D{.}{.}{-3}} 
\\[-1.8ex]\hline 
\hline  
 & \multicolumn{7}{c}{\textit{Dependent variable:}} \\ 
\cline{2-8} 
 & \multicolumn{7}{c}{GradePointUnit} \\ 
 & \multicolumn{1}{c}{\textit{generalized}} & \multicolumn{5}{c}{\textit{linear}} \\[-1.8ex]
 & \multicolumn{1}{c}{\textit{least squares}} & \multicolumn{5}{c}{\textit{mixed effects}} \\ 
 & \multicolumn{1}{c}{(1)} & \multicolumn{1}{c}{(2)} & \multicolumn{1}{c}{(3)} & \multicolumn{1}{c}{(4)} & \multicolumn{1}{c}{(5)} & \multicolumn{1}{c}{(6)} & \multicolumn{1}{c}{Random Effects SD}\\ 
\hline  
 GPE  &  &  &  & 0.436^{***} & 0.444^{***} & 0.502^{***} & 0.042  \\ 
  &  &  &  & (0.010) & (0.010) & (0.017) \\ 
  & & & & & & \\[-5ex] 
 GenderM &  &  &  &  & -0.418^{***} & -0.075 &  0.461\\ 
  &  &  &  &  & (0.049) & (0.122) \\ 
  & & & & & & \\[-5ex] 
 Constant & 4.002^{***} & 3.578^{***} & 3.604^{***} & 3.683^{***} & 3.794^{***} & 3.737^{***} & 0.713 \\ 
  & (0.025) & (0.154) & (0.136) & (0.144) & (0.148) & (0.166) \\ 
  & & & & & & & Residual: 2.758\\ 
\hline  
Observations & \multicolumn{1}{c}{16,102} & \multicolumn{1}{c}{16,102} & \multicolumn{1}{c}{16,102} & \multicolumn{1}{c}{16,102} & \multicolumn{1}{c}{16,102} & \multicolumn{1}{c}{16,102} \\ 
Log Likelihood & \multicolumn{1}{c}{-41,374} & \multicolumn{1}{c}{-40,821} & \multicolumn{1}{c}{-40,420} & \multicolumn{1}{c}{-39,485} & \multicolumn{1}{c}{-39,449} & \multicolumn{1}{c}{-39,352} \\ 
Akaike Inf. Crit. & \multicolumn{1}{c}{82,753} & \multicolumn{1}{c}{81,649} & \multicolumn{1}{c}{80,847} & \multicolumn{1}{c}{78,979} & \multicolumn{1}{c}{78,910} & \multicolumn{1}{c}{78,737} \\ 
Bayesian Inf. Crit. & \multicolumn{1}{c}{82,768} & \multicolumn{1}{c}{81,672} & \multicolumn{1}{c}{80,878} & \multicolumn{1}{c}{79,018} & \multicolumn{1}{c}{78,956} & \multicolumn{1}{c}{78,860} \\ 
\hline 
\hline  
\textit{Note:}  & \multicolumn{6}{r}{$^{*}$p$<$0.1; $^{**}$p$<$0.05; $^{***}$p$<$0.01} \\ 
\end{tabular} 
\caption{\label{tab:RQ2_1}Table summing up a multilevel model, where our final model (6) consists of 3 levels. used to predict grades in first year and second year physics courses. (1) shows the the baseline model showing the intercept only. (2) shows the same model, but with the student's grades nested within each physics course. (3) shows the model with the term (i.e., year and semester) a student took the course in added as another level. This represents the structure of our final models, where student's grade is nested within a specific course, and in a specific term. (4) shows a fixed slopes, random intercepts model with GPE added as a predictor variable. (5) shows the model where GPE and gender are added as predictor variables. (6) shows the same model as (5), but with both random intercepts and random slopes. As can be seen from model (6), overall there were not significant gender differences in course grade, after controlling for GPE. However the random effects standard deviation shows that there was considerable variation in gender differences between courses.} 
\end{table*} 
\end{turnpage}

\begin{figure*}[hp]
\centering
\includegraphics[width=15cm, height = 20cm]{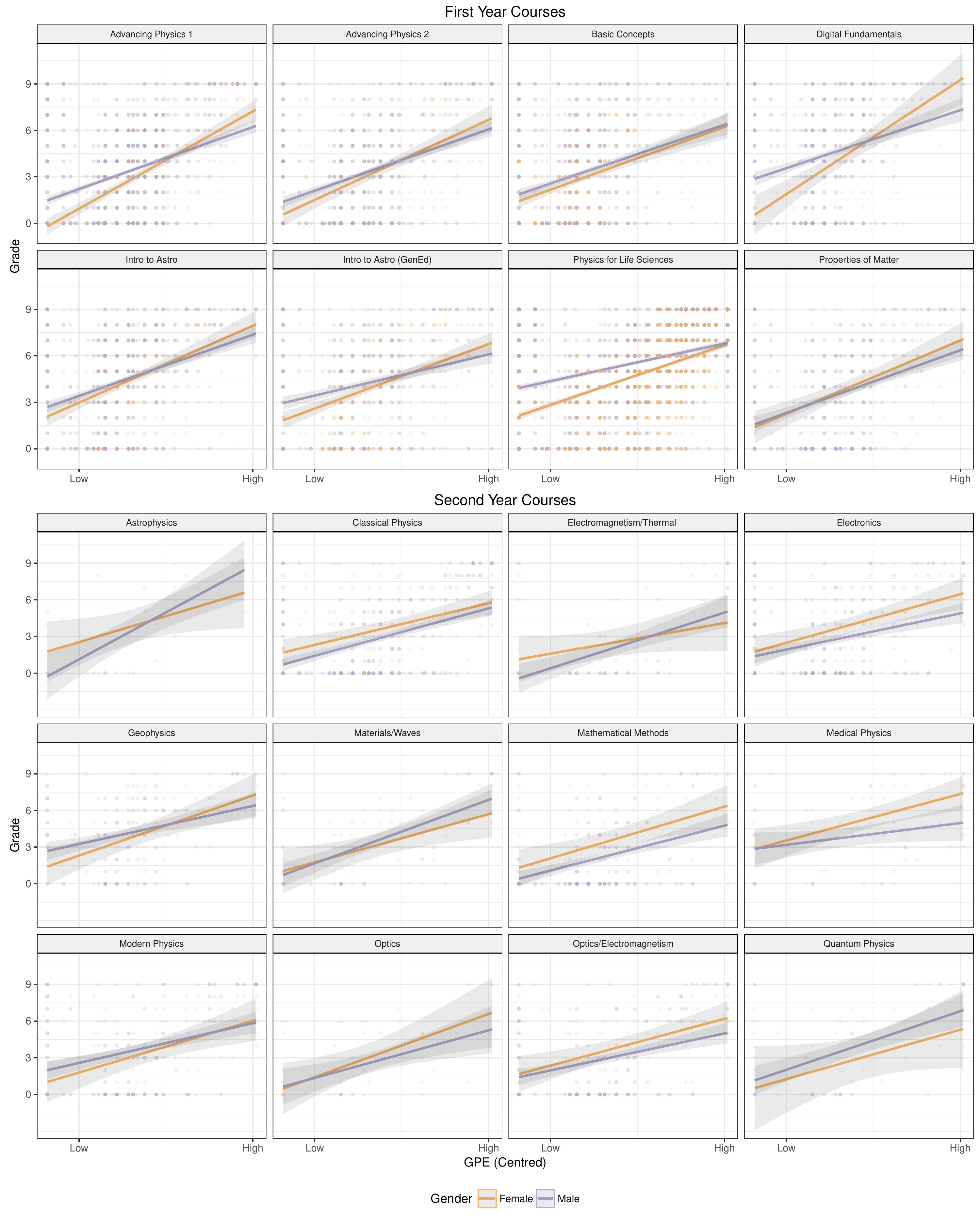}
\caption{\label{fig:MLM_LinearModels} The above plots show linear regression models with GPE predicting grades in key first and second year physics courses, as a function of gender. As outlined by the multilevel model presented in \ref{tab:RQ2_1}, gender differences varied between courses, although certain trends can be made out. In first year courses where there were significant gender differences for students studying Advancing Physics 1, Digital Fundamentals, Intro to Astro (General Education) and Physics for Life Sciences. In these courses, male students with low GPE scores tended to get high course grades than their female counterparts. In second year courses, there did not appear to be statistically significant gender differences - although, if anything, the gender differences in course performance may have favored female students. However, we cannot be confident in any gender differences in second year courses; it is likely that lower course populations in second year may have contributed to reduced statistical confidence.}
\end{figure*}

\begin{table*}[!htbp] \centering 

\begin{tabular}{@{\extracolsep{5pt}}lc} 
\\\hline 
\hline 
 & \multicolumn{1}{c}{\textit{Dependent variable:}} \\ 
\cline{2-2} 
\\ & GradePointUnit \\ 
\hline 
 GPE (Centered) & 1.890$^{***}$ \\ 
  & (0.205) \\ 
  & \\[-3ex] 
 Gender & $-$0.307 \\ 
  & (0.270) \\ 
  & \\ [-3ex] 
 High School Type & 0.411$^{**}$ \\ 
  & (0.198) \\ 
  & \\[-3ex]  
 GPE (Centered):Gender & 0.035 \\ 
  & (0.260) \\ 
  & \\[-3ex]  
 GPE (Centered):High School Type & $-$0.535$^{**}$ \\ 
  & (0.218) \\ 
  & \\[-3ex]  
 Gender:High School Type & $-$0.187 \\ 
  & (0.333) \\ 
  & \\[-3ex]  
 GPE (Centered):Gender:High School Type & 0.135 \\ 
  & (0.317) \\ 
  & \\[-3ex]  
 Constant & 2.957$^{***}$ \\ 
  & (0.184) \\ 
\hline  
Observations & 1,785 \\ 
R$^{2}$ & 0.269 \\ 
Adjusted R$^{2}$ & 0.266 \\ 
Residual Std. Error & 2.491 (df = 1777) \\ 
F Statistic & 93.535$^{***}$ (df = 7; 1777) \\ 
\hline 
\hline \\ 
\textit{Note:}  & \multicolumn{1}{r}{$^{*}$p$<$0.1; $^{**}$p$<$0.05; $^{***}$p$<$0.01} \\ 
\end{tabular} 
\caption{\label{tab:RQ2_LM}Regression output for a 3 way interaction between GPE (centered), gender and high~school~type predicting AP1 grade. GPE, gender, high school type, and a GPE$\times$high~school~type interaction were all significant predictors of AP1 grade. This shows that individuals who were academically prepared (based on GPE) were more likely to do well in AP1. Individuals who attended single-sex schools were more likely to do well in AP1, but only if they had high GPE scores. Individuals from single-sex schools who were not academically prepared tended to do less well in AP1 than students from co-educational schools.}
\end{table*} 

\begin{figure*}[hp]
\centering
\includegraphics[width=15cm, height = 10cm]{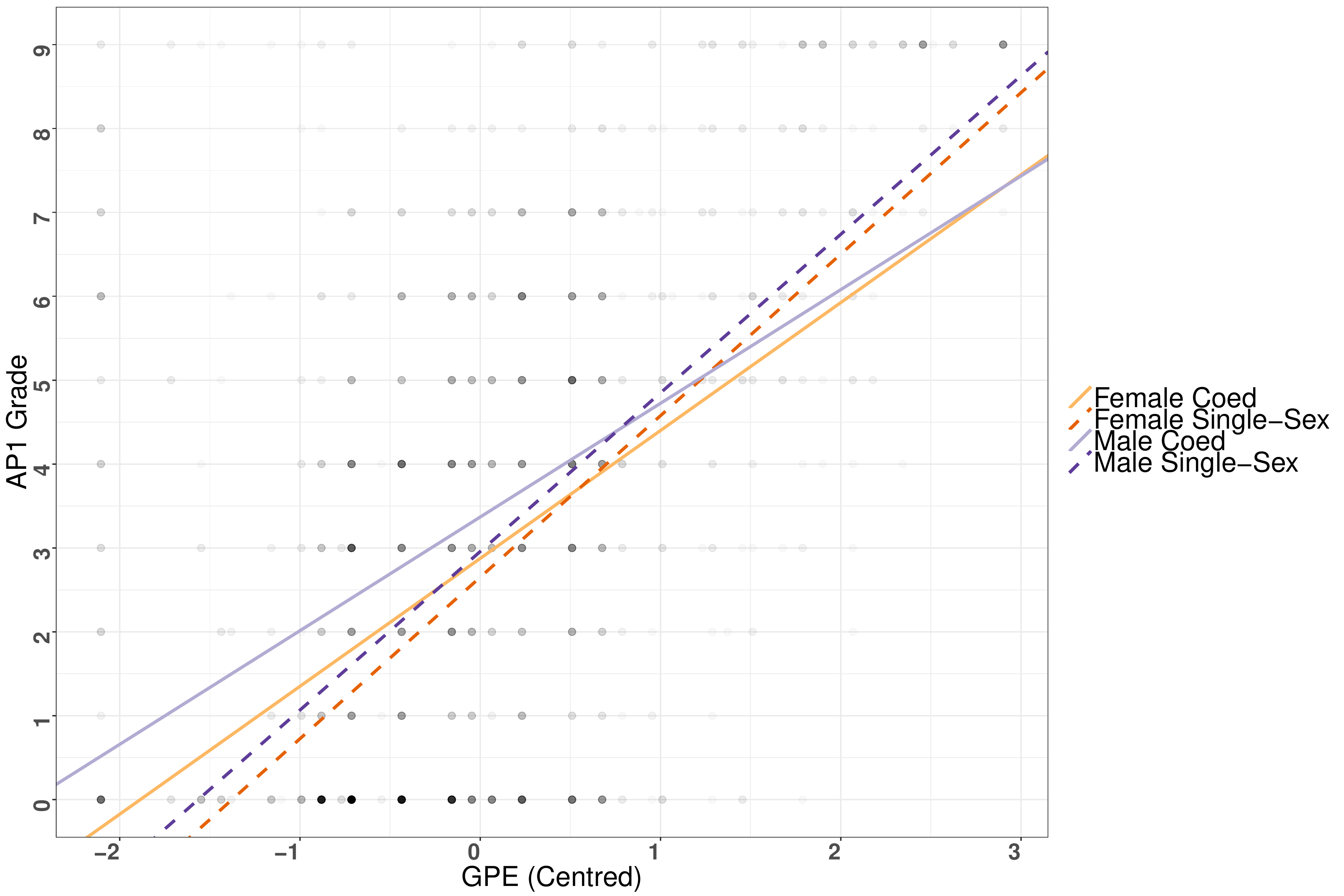}
\caption{\label{fig:A}Scatter plot visualising the results of the linear regression model shown in Table \ref{tab:RQ2_LM}, with GPE (mean centered), gender and high~school~type predicting AP1 grade~point~unit. At low GPEs, male students and those from co-educational school outperformed those from other groups. At high GPEs, students from single-sex schools outperformed those from co-educational schools, while there was no gender difference. Plot point transparency is dependent on number of students.}
\end{figure*}

\begin{table*}[bhpt]
\centering
\begin{tabular}{ r|c|c| }
\multicolumn{1}{r}{}
 &  \multicolumn{1}{c}{\parbox{4cm}{\begin{center} Took Basic Concepts of Physics \\ before AP1\\\end{center}}}
 & \multicolumn{1}{c}{\parbox{4cm}{\begin{center}
  Did not take Basic Concepts of Physics \\before AP1\\ \end{center}}} \\
\cline{2-3}
Male & 5 & 623 \\
\cline{2-3}
Female & 7 & 175 \\
\cline{2-3}

\end{tabular}
\caption{\label{tab:RQ3}The number of academically prepared students (GPE $>$ 3) that took Basic Concepts of Physics before taking AP1 as a function of gender. Given no gender differences, it would be expected that two, rather than seven, female students, and nine, rather than five, male students would have taken Basic Concepts of Physics beforehand.}
\end{table*}

\begin{table*}[bhpt]
\centering
\begin{tabular}{r|c|c| }
\multicolumn{1}{r}{}
 &  \multicolumn{1}{c}{\parbox{4cm}{\begin{center} Took Basic Concepts of Physics \\ before AP1\\\end{center}}}
 & \multicolumn{1}{c}{\parbox{4cm}{\begin{center}
  Did not take Basic Concepts of Physics \\before AP1\\ \end{center}}} \\
\cline{2-3}
Male & 56 & 429 \\
\cline{2-3}
Female & 23 & 101 \\
\cline{2-3}

\end{tabular}
\caption{\label{tab:RQ3_2}The number of academically unprepared students (GPE $\leq$ 3) that took Basic Concepts of Physics before taking AP1 as a function of gender. Given no gender differences, it would be expected that 15, rather than  23, female students, and 64, rather than 56, male students would have taken Basic Concepts of Physics beforehand.}
\end{table*}

\begin{table*}[h]
\centering
\begin{tabular}{ |c|c| }
  \hline
  GPE & Odds Ratio (SE of log odds) \\ \hline
   $>$ 3 & 4.98 (0.59)** \\ \hline
   $\leq$ 3 & 1.74 (0.27)*\\ 
  
  \hline
\end{tabular}
\caption{\label{tab:RQ3_3}The odds of a female student taking Basic Concepts of Physics before AP1 over a male student. Female students were more likely to take Basic Concepts of Physics before AP1, regardless of academic preparation. $^{*}$p$<$ .1; $^{**}$p$<$ .05}
\end{table*}


\end{document}